\documentstyle[aps]{revtex}

\begin{document}
\title{Entropy gap and time asymmetry II. }
\author{Roberto Aquilano}
\address{Instituto de F\'\i sica Rosario,\\
Bvd. 27 de Febrero 210bis, 2000 Rosario, Argentina, and\\
Observatorio Astron\'{o}mico Municipal Rosario,\\
Casilla de Correos 606, 2000 Rosario, Argentina.}
\author{Mario Castagnino, Ernesto Eiroa}
\address{Instituto de Astronom\'{\i}a y F\'{\i}sica del Espacio.\\
Casilla de Correos 67, Sucursal 28, 1428 Buenos Aires, Argentina.}
\maketitle

\begin{abstract}
In this letter the paper [R. Aquilano, M. Castagnino, Mod. Phys. Lett A, 
{\bf 11}, 755 (1996)] is improved by considering that the main source of
entropy production are the stars photospheres.
\end{abstract}

\section{Introduction.}

In papers\cite{I} \footnote{%
It is a good occasion to correct some errata in the abstract of paper \cite
{I}.
\par
-In line 2: it reads ''quantitative'', it must read ''qualitative''.
\par
-In line 3: it reads ''entropy'', it must read ''entropy gap''.
\par
-In line 4: it reads ''qualitative'', it must read ''quantitative''.} and 
\cite{Cast} two of us reported a rough coincidence between the time where
the minimum of the entropy gap $\Delta S=S_{act}-S_{\max }$ \cite{Davies},
takes place and the time where all the stars will exhaust their fuel. The
rough nature of this result was explained saying that the calculation was
done using a model extremely naive and simplified: a homogeneous universe 
\footnote{%
Besides that the higher order terms in eq. (3) of ref. \cite{I} were
neglected; perhaps they may be important for finite times.}. In fact in the
real universe nuclear reactions (that were considered as the main source of
entropy) take place within the stars, that can only be properly considered
in an inhomogeneous geometry. So we expend some time trying to generalize
our model to these type of geometries and computing the entropy of the
galaxies and stars in the condensation period, after decoupling time, in
order to make the coincidence precise \footnote{%
In fact, inhomogeneity must be considered, but in a different way, as we
will see below.}. It was soon clear to us that the latter entropy can be
neglected with respect to the amount of entropy produced within the stars
and therefore the mechanism that produced the inhomogeneity is not so
important for our problem. Stars are formed when a sufficiently large mass
of interstellar gas is compressed into a small enough volume, so its force
of self-gravitation becomes sufficiently great to cause gravitational
collapse. The instability makes the cloud to reduce its size quickly, the
temperature rises and pressure forces begin to restrict the collapse until
hydrostatic equilibrium is obtained. In this process, gravitational energy
is converted in kinetic energy and radiation\cite{Clayton}. In the classical
model of star formation the time of contraction is the Kelvin-Helmholtz time%
\cite{protostars}:

\begin{equation}
t_{KH}=\frac{GM^2}{RL}  \label{12}
\end{equation}

where $M$ is the mass, $R$ is the radius and $L$ is the luminosity of the
star. For a star like the sun this time is $3$ $10^7$ years \cite{Clayton} 
\cite{protostars}. In more recent models of star formation, the relevant
time scale is the much shorter free-fall time \cite{protostars}:

\begin{equation}
t_{ff}=(\rho G)^{-1/2}=7\cdot 10^5\left( \frac n{10^4}\right) ^{-1/2}years
\label{13}
\end{equation}

where $\rho $ and $n$ are the cloud mass density and the cloud number
density respectively.

In any case, it takes less than $1\%$ of the lifetime $(\cong Gyr)$ to the
star to form. In the contraction to make the star, the energy radiated is a
part of the gravitational energy and the nuclear reactions are not
important. The binding gravitational energy is negligible compared with the
energy liberated through nuclear reactions when the star is already formed 
\cite{Clayton}. As a consequence we can neglect the entropy production in
the phase of contraction that makes the star.

So, in order to improve our rough result, we were forced to change the
approach.

Let us review the main formulae. The entropy gap was the conditional entropy
of $\rho (t)$, the state of the universe at time $t$ with respect to the
equilibrium state at that time $\rho _{*}(t)$ :

\begin{equation}
\Delta S=-tr[\rho \log (\rho _{*}^{-1}\rho )]  \label{A}
\end{equation}

where

\begin{equation}
\rho (t)=\rho _{*}(t)+\rho _1\exp (-\gamma t)+O[(\gamma t)^{-1}]  \label{B}
\end{equation}

and

\begin{equation}
\rho _{*}(\omega )=ZT^{-3}\frac 1{\exp (\frac \omega T-1)}  \label{C}
\end{equation}

$\rho _1$ was a phenomenological coefficient constant in time, the time
variation of the main irreversible process was $\exp (-\gamma t)$ so $\gamma
^{-1}$ was the characteristic time of this process, $T$ was the temperature
of the universe, $T\sim a^{-1}$ where $a$ is the radius or scale of the
universe, and $Z$ is a normalization constant.

Then we could approximate $\Delta S$ as 
\begin{equation}
\Delta S=-CT^3\exp (-\gamma t)\exp (\frac{\omega _1}T)  \label{D}
\end{equation}

where $\omega _1$ was the characteristic energy where $\rho _1$ is peaked.

The time where the minimum of $\Delta S$ is located was: 
\begin{equation}
t_{cr}\approx t_0\left( \frac 23\frac{\omega _1}{T_0}\frac{t_{NR}}{t_0}%
\right) ^3  \label{14}
\end{equation}

We selected numerical values for the parameters: $\omega _1=T_{NR},$ the
temperature of the nuclear reactions within the stars (that was used in
paper \cite{I} as the main source of entropy), $t_{NR}=\gamma ^{-1}$ the
characteristic time of these nuclear reactions, $t_0$ the age of the
universe, and $T_0,$ the cosmic micro-wave background temperature.

$T_{NR}$ and $t_{NR}$ were chosen between the following values \cite{Cumul}: 
\begin{equation}
T_{NR}=10^6\text{ }to\text{ }10^8{}\text{ }K  \label{15}
\end{equation}
\[
t_{NR}=10^6\text{ }to\text{ }10^9\text{ }years 
\]
while for $t_0$ and $T_0$ we can take: 
\begin{equation}
t_0=1.5\times 10^{10}\text{ }years  \label{16}
\end{equation}
\[
T_0=3K 
\]
In order to obtain a reasonable result we chose (with no explanation) the
lower bounds for $T_{NR}$ and $t_{NR\text{ }}$ and for $t_{cr}$ we obtained
: 
\begin{equation}
t_{cr}\preceq 10^4t_0\cong 10^{14}years  \label{18}
\end{equation}

concluding that the order of magnitude of $t_{cr}$ was a realistic one. In
fact, $10^4t_0\approx 1.5\times 10^{14}years$ after the big-bang all the
stars will exhaust their fuel \cite{AJP}\footnote{%
After the publication of paper \cite{I}, in 1996, new data about this time
appear in \cite{LaBo}.}, so it is reasonable that this time would be of the
same order than the one where the entropy gap stops its decreasing and
begins to grow \cite{Reeves}. But the choice of the lower bound in eqs. (\ref
{15}) was not explained and only the inhomogeneity was argued as above.

\section{The photosphere as the main unstable system.}

Now we have reconsidered the problem and we conclude that, even if nuclear
reactions within the stars are the main source of entropy, the parameters $%
T_{NR}$ and $t_{NR}$ are not the good ones to define the behavior of the
term $\exp (-\gamma t)\rho _1$ of equation (\ref{A})((3) of paper \cite{I}),
since they do not correspond to the main unstable system that we must
consider \footnote{%
In fact, to take the parameters $T_{NR}$ and $t_{NR}$ corresponds to take an
homogenous gas model that fills the universe where, nuclear reactions take
place in this homogenous gas, while in real universe these reactions take
place within the star in a quite inhomogeneous scenario.}. In fact the main
production of entropy in a star is not located in its core, where the
temperature is almost constant (and equal to $T_{NR})$, but in the
photosphere where the star radiates. The energy radiated from the surface of
the star is produced in the interior by fusion of light nuclei into heavier
nuclei. Most stellar structures are essentially static, so the power
radiated is supplied at the same rate by these exothermic nuclear reactions
that take place near the center of the star \cite{Clayton}. Once the star is
formed, it settles into a termally stable state where all the nuclear energy
is radiated at the surface and the rate of internal entropy change is
extremely low \cite{protostars}.

We can decompose the whole star in two branch systems\cite{Davies}, as
explained in paper \cite{I} (or in section VII of paper \cite{Cast}), where
a chain of branch systems was introduced. We have two branch systems to
study: the core and the photosphere. The core gives energy to the
photosphere and in turn the photosphere diffuses this energy to the
surroundings of the star, namely in the bath of microwave radiation at
temperature $T_{0\text{ .}}$In this way, we have two sources of entropy
production: the radiation of energy at the surface of the star and the
change of composition inside the star (as time passes we have more helium
and less hydrogen). Since the core of a star is near thermodynamic
equilibrium, we neglect the second and we concentrate on the first: the
radiation from the surface of the star (related with the difference between
the star and the background temperatures). So the temperature of the
photosphere and not the one of the core must be introduced in our formula.
This is also the case for the lifetime. We must take the lifetime of the
photosphere not the one of the nuclear reactions. Thus it is better to
consider the photosphere as the unstable system that defines the term $\exp
(-\gamma t)\rho _1$ of equation (\ref{B}). So we must change $T_{NR}$ and $%
t_{NR}$ by $T_P$, the temperature of the photosphere and $t_S$ the
characteristic lifetime of the star respectively. Then we must change eq. (%
\ref{14}) to: 
\begin{equation}
t_{cr}\approx t_0\left( \frac 23\frac{T_P}{T_0}\frac{t_S}{t_0}\right) ^3
\label{19}
\end{equation}

Considering that the mean mass of stars is $0.64M_{\odot }$ with surface
temperature $4.6\cdot 10^3K$ and lifetime $3.8\cdot 10^{10}years$ (see
appendix) we obtain:

\begin{equation}
t_{cr}\cong 10^{10}t_0\cong 10^{20}years  \label{20}
\end{equation}

but now the computation was not done using an arbitrary choice of the lower
bound in some data, but using the first meaningful figure \footnote{%
In paper \cite{PRD}, using the photosphere data, but just orders of
magnitude, we have again obtained (\ref{18}).} in all the data \footnote{%
We have obtained a larger result than (\ref{18}), but remember that the
latter was obtained just by taking the lower bound in eqs. (\ref{15}). If we
would choose the mean values in these equations we would obtain a larger
result than the one of eq. (\ref{20}), without any significance.}.

\section{The stelliferous and the degenerated eras.}

Let us now compare this result with those of paper \cite{AdLa}, where the
future history of the universe is analyzed. $t_{cr}\cong
10^{10}t_0=10^{20}years$ is placed after the end of the ''stelliferous era''
($10^6<t<10^{14}years)$, where most of the energy generated in the universe
arises from nuclear processes in the conventional star evolution, and at the
beginning of the ''degenerated era'' ($10^{15}<t<10^{37}years)$, where most
of the (baryonic) mass of the universe is locked up in degenerated stellar
objects: brown dwarfs, white dwarfs, and neutron stars. In this era energy
is generated through proton decay and particle annihilation.

So $t_{cr}\cong 10^{20}years$ is not a bad place for the minimum of the
entropy gap, since it is bigger than the end of conventional star formation (%
$10^{14}years)$, it is also bigger than the typical time of star formation
via brown dwarfs collision ($10^{16}years)$, and it is of the order of the
time corresponding to stellar evaporation from galaxies ($10^{19}years)$.
Namely a time where we can consider that the main mechanisms of formation of
stars are ended while the beginning of the evaporation process has just
started. This is the best place for $t_{cr}$, that must be the frontier
between the growing order period (formation of structures with decaying
entropy \cite{Reeves}) and the diminishing order period (decaying of the
structures with growing entropy). Tacking into account the great uncertainty
of all cosmological data we can say that $t_{cr}$ is located in the edge
between the stelliferous normal era, where we are living and which is
dominated by the formation of structures, and the future degenerated era,
full of, by now strange, objects, where the growing of disorder will begin.
So even if eq. (\ref{19}) is the result of a very simple model it gives a
very reasonable value.

\section{Conclusion.}

By choosing a more realistic model as the main source of entropy production,
the photosphere of the stars, we have obtained a reasonable value for the
time where the minimum of the entropy gap is reached. This is the frontier
of the formation and destruction of structure periods and therefore one of
the most important moments of the future history of the Universe.

\section{Acknowledgments.}

We wish to thank Omar Benvenutto for fruitful discussions. This work was
partially supported by grants Nos. CI1*-CT94-0004 of the European Community,
PID-0150 and PEI-0126-97 of CONICET (National Research Council of
Argentina), EX-198 of the Buenos Aires University and 12217/1 of
Fundaci\'{o}n Antorchas and the British Council.

\section{Appendix}

The stars have masses in the interval \cite{Uns},\cite{AdFa}: 
\begin{equation}
0.1\leq m=\frac{M_{*}}{M_{\odot }}\leq 100  \label{A0}
\end{equation}

where $M_{*}$ is the mass of the star and $M_{\odot }$ is the mass of the
Sun.

The mean mass of stars can be calculated using the initial mass function
(IMF) $\xi (m)$ defined by \cite{AdFa}:

\begin{equation}
dN=\xi (m)\cdot dm  \label{A1}
\end{equation}
where $dN$ is the number of stars with masses between $m$ and $m+dm$

We will use as IMF \cite{AdLa}:

\begin{equation}
\frac{dN}{d(\ln m)}=\psi (\ln m)=\exp \left[ A-\frac 1{2<\sigma >^2}\ln
^2\left( \frac m{m_c}\right) \right]  \label{A2}
\end{equation}

where $<\sigma >\cong 1.57$ , $m_c\cong 0.1$ and $A$ is a constant that sets
the overall normalization of the distribution (which is not important
because it cancels in the calculation of the mean mass of stars). This form
of the IMF is consistent with observations (see \cite{MiSca}).As

\begin{equation}
\xi (m)=\psi (\ln m)\cdot \frac{d(\ln m)}{dm}=\frac 1m\psi (\ln m)=\frac 1m%
\exp \left[ A-\frac 1{2<\sigma >^2}\ln ^2\left( \frac m{m_c}\right) \right]
\label{A3}
\end{equation}

then the mean mass of stars is:

\begin{equation}
<m>=\frac{\int\limits_{0.1}^{100}m\cdot \xi (m)\cdot dm}{\int%
\limits_{0.1}^{100}\xi (m)\cdot dm}=0.64  \label{A4}
\end{equation}

Let us calculate the effective surface temperature of a star with mass $%
m=0.64$. The Luminosity of a star is given by the Stefan- Boltzmann law:

\begin{equation}
L=4\pi R^2\sigma T_p^4  \label{A5}
\end{equation}

where $R$ is the radius of the star, $\sigma $ is the Stefan- Boltzmann
constant and $T_p$ is the effective temperature of the surface of the star.
Then

\begin{equation}
\frac L{L_{\odot }}=\left( \frac R{R_{\odot }}\right) ^2\left( \frac{T_p}{%
T_{\odot }}\right) ^4  \label{A6}
\end{equation}

where $L_{\odot }$ , $R_{\odot }$ ,$T_{\odot }$ are, respectively, the
luminosity, the radius and the surface temperature of the Sun.

The radius and luminosity for stars in the main sequence, as a function of
mass are (see \cite{Kipp}):

\begin{equation}
\frac R{R_{\odot }}\cong \left( \frac{M_{*}}{M_{\odot }}\right) ^\beta
=m^\beta  \label{A7}
\end{equation}

\begin{equation}
\frac L{L_{\odot }}\cong \left( \frac{M_{*}}{M_{\odot }}\right) ^\eta =m^\eta
\label{A8}
\end{equation}

with $\beta \cong 0.6$ for stars of low mass and $\eta \cong 3.2$ .So we have

\begin{equation}
T_P\cong T_{\odot }\left( \frac{M_{*}}{M_{\odot }}\right) ^{\frac{\eta
-2\beta }4}\cong T_{\odot }\cdot m^{0.5}  \label{A9}
\end{equation}

Using that $T_{\odot }\cong 5780K$ and $m=0.64$ we obtain

\begin{equation}
T_P\cong 4.6\cdot 10^3K  \label{A10}
\end{equation}

The lifetime of a star can be calculated by the equation(see \cite{AdLa} ):

\begin{equation}
t_s=10^{10}\left( \frac{M_{*}}{M_{\odot }}\right) ^{-\alpha
}years=10^{10}m^{-\alpha }years  \label{A11}
\end{equation}

where $\alpha \cong 3-4$ for stars of low mass. Taking $\alpha =3$ we have:

\begin{equation}
t_s\cong 3.8\cdot 10^{10}years  \label{A12}
\end{equation}

\end{document}